\documentclass[]{ws-ijmpa}
\usepackage{amsmath}
\usepackage{amssymb}

\begin{document}

\title{Exotic smooth $\mathbb{R}^{4}$ and certain configurations of NS
and D branes in string theory%
\thanks{Based on the talk ,,Small exotic smooth $\mathbb{R}^{4}$ and string
theory'' given at the International Congress of Mathematicians, ICM2010,
19-28.08.2010, Hyderabad, India%
} }

\author{Torsten Asselmeyer-Maluga}%
\address{German Aero space center, Rutherfordstr. 2, 12489 Berlin \\ torsten.asselmeyer-maluga@dlr.de%
} 
\author{Jerzy Kr\'ol}%
\address{University of Silesia, Institute of Physics, ul. Uniwesytecka 4, 40-007
Katowice \\ iriking@wp.pl %
}

\maketitle
\begin{history}
\received{Day Month Year}
\revised{Day Month Year}
\end{history}

\begin{abstract}
In this paper we show that in some important cases 4-dimensional data
can be extracted from superstring theory such that a) the data are
4 Euclidean geometries embedded in standard $\mathbb{R}^{4}$, b)
these data depend on NS and D brane charges of some string backgrounds,
c) it is of potential relevance to 4-dimensional physics, d) the compactification
and stabilization techniques are not in use, but rather are replaced.
We analyze certain configurations of NS and D-branes in the context
of $SU(2)$ WZW model and find the correlations with different exotic
smoothings of $\mathbb{R}^{4}$. First, the dynamics of D-branes in
$SU(2)$ WZW model at finite $k$, i.e. the charges of the branes,
refers to the exoticness of ambient $\mathbb{R}^{4}$. Next, the correspondence
between exotic smoothness on 4-space, transversal to the world volume
of NS5 branes in IIA type, and the number of these NS5 branes follows.
Finally, the translation of 10 dimensional string backgrounds to 4
Euclidean spaces embedded as open subsets in the standard $\mathbb{R}^{4}$
is achieved. 

\end{abstract}
%\tableofcontents{}

\ccode{PACS numbers: 11.25.Uv, 02.40K, 04.20Gz}

\section{Introduction}

The problem with successful inclusion of effects of exotic open 4-manifolds
like exotic $\mathbb{R}^{4}$ into any physical theory, is the notorious
lack of an explicit coordinate-like presentation of these smooth manifolds.
In the series of our recent papers we addressed this issue and worked
out some relative techniques allowing for analytical treatment of
small exotic $\mathbb{R}^{4}$'s \cite{AsselmeyerKrol2009,AsselmeyerKrol2009a,AsselmeyerKrol2010,Krol2010}.
Based on these results we will show in this paper a rather unexpected
relation between configurations of D-branes in some exact string backgrounds
with the exotic smoothness structure of the $\mathbb{R}^{4}$. This
relation is not only a pure formal correspondence but instead we see
it as a way to 4-dimensional physics. The proposed realization is
different and independent on various compactifications or model building
techniques worked out so far by string theorists and aiming also at
the description of our 4-dimensional world. Why do we develop any
alternative to the well established phenomenological approach in string
theory? The answer is rather direct. First, in spite of the substantial
effort results worked out in string theory are highly ambiguous: there
exist about $10^{500}$ possible backgrounds as candidates for real
physics. Second, the appearance of exotic smoothness of Euclidean
$\mathbb{R}^{4}$ in the formalism of string theory is a direct indication
that this formalism deals with dimension 4 at the fundamental level.
Even from a very general point of view, if different string constructions
refer indeed to exotic 4-smoothness on Euclidean 4-space, there is
a chance to reduce the above-mentioned ambiguity. When additionally
we would have a good understanding of how 4-exotics refer to real
physics there is a big chance to solve the puzzle of a relation between
string theory and 4-dimensional physics. Even though much remain to
be done this paper serves as the first important step into this direction. 

The basic technical ingredient of the analysis of small exotic $\mathbb{R}^{4}$'s
enabling uncovering many applications also in string theory is the
relation between exotic (small) $\mathbb{R}^{4}$'s and non-cobordant
codimension-1 foliations of $S^{3}$ as well gropes and wild embeddings
as shown in \cite{AsselmeyerKrol2009}. The foliations are classified
by the Godbillon-Vey classes as elements of the cohomology group $H^{3}(S^{3},\mathbb{R})$.
By using $S^{1}$-gerbes it was possible to interpret the integral
elements $H^{3}(S^{3},\mathbb{Z})$ as characteristic classes of the
$S^{1}$-gerbes over $S^{3}$ \cite{AsselmeyerKrol2009a}. In the
next section we will explain the whole complex of ideas more carefully.
The following section deal with the relation between string backgrounds
and exotic $\mathbb{R}^{4}$. A discussion of the results closes the
paper.

\section{Exotic $\mathbb{R}^{4}$ and codimension-one foliations of the 3-sphere}

The main line of the topological argumentation can be briefly described
as follows:

\begin{enumerate}
\item In Bizaca's exotic $\mathbb{R}^{4}$ one starts with the neighborhood
$N(A)$ of the Akbulut cork $A$ in the K3 surface $M$. The exotic
$\mathbb{R}^{4}$ is the interior of $N(A)$.
\item This neighborhood $N(A)$ decomposes into $A$ and a Casson handle
representing the non-trivial involution of the cork.
\item From the Casson handle we construct a grope containing Alexanders
horned sphere.
\item Akbulut's construction gives a non-trivial involution, i.e. the double
of that construction is the identity map.
\item From the grope we get a polygon in the hyperbolic space $\mathbb{H}^{2}$.
\item This polygon defines a codimension-1 foliation of the 3-sphere inside
of the exotic $\mathbb{R}^{4}$ with an wildly embedded 2-sphere,
Alexanders horned sphere.
\item Finally we get a relation between codimension-1 foliations of the
3-sphere and exotic $\mathbb{R}^{4}$.
\end{enumerate}
Now we will explain the details in this construction.

An exotic $\mathbb{R}^{4}$ is a topological space with $\mathbb{R}^{4}-$topology
but with a different (i.e. non-diffeomorphic) smoothness structure
than the standard $\mathbb{R}_{std}^{4}$ getting its differential
structure from the product $\mathbb{R}\times\mathbb{R}\times\mathbb{R}\times\mathbb{R}$.
The exotic $\mathbb{R}^{4}$ is the only Euclidean space $\mathbb{R}^{n}$
with an exotic smoothness structure. The exotic $\mathbb{R}^{4}$
can be constructed in two ways: by the failure to arbitrarily split
a smooth 4-manifold into pieces (large exotic $\mathbb{R}^{4}$) and
by the failure of the so-called smooth h-cobordism theorem (small
exotic $\mathbb{R}^{4}$). Here we will use the second method. 

Consider the following situation: one has two topologically equivalent
(i.e. homeomorphic), simply-connected, smooth 4-manifolds $M,M'$,
which are not diffeomorphic. There are two ways to compare them. First
one calculates differential-topological invariants like Donaldson
polynomials \cite{DonKro:90} or Seiberg-Witten invariants \cite{Akb:96}.
But there is another possibility: It is known that one can change
a manifold $M$ to $M'$ by using a series of operations called surgeries.
This procedure can be visualized by a 5-manifold $W$, the cobordism.
The cobordism $W$ is a 5-manifold having the boundary $\partial W=M\sqcup M'$.
If the embedding of both manifolds $M,M'$ in to $W$ induces homotopy-equivalences
then $W$ is called an h-cobordism. Furthermore we assume that both
manifolds $M,M'$ are compact, closed (no boundary) and simply-connected.
As Freedman \cite{Fre:82} showed a h cobordism implies a homeomorphism,
i.e. h-cobordant and homeomorphic are equivalent relations in that
case. Furthermore, for that case the mathematicians \cite{CuFrHsSt:97}
are able to prove a structure theorem for such h-cobordisms:\\
 \emph{Let $W$ be a h-cobordism between $M,M'$. Then there are contractable
submanifolds $A\subset M,A'\subset M'$ together with a sub-cobordism
$V\subset W$ with $\partial V=A\sqcup A'$, so that the h-cobordism
$W\setminus V$ induces a diffeomorphism between $M\setminus A$ and
$M'\setminus A'$.} \\
 Thus, the smoothness of $M$ is completely determined (see also \cite{Akbulut08,Akbulut09})
by the contractible submanifold $A$ and its embedding $A\hookrightarrow M$
determined by a map $\tau:\partial A\to\partial A$ with $\tau\circ\tau=id_{\partial A}$
and $\tau\not=\pm id_{\partial A}$($\tau$ is an involution). One
calls $A$, the \emph{Akbulut cork}. According to Freedman \cite{Fre:82},
the boundary of every contractible 4-manifold is a homology 3-sphere.
This theorem was used to construct an exotic $\mathbb{R}^{4}$. Then
one considers a tubular neighborhood of the sub-cobordism $V$ between
$A$ and $A'$. The interior $int(V)$ (as open manifold) of $V$
is homeomorphic to $\mathbb{R}^{4}$. If (and only if) $M$ and $M'$
are homeomorphic, but non-diffeomorphic 4-manifolds then $int(V)\cap M$
is an exotic $\mathbb{R}^{4}$. As shown by Bizaca and Gompf \cite{Biz:94a,BizGom:96}
one can use $int(V)$ to construct an explicit handle decomposition
of the exotic $\mathbb{R}^{4}$. We refer for the details of the construction
to the papers or to the book \cite{GomSti:1999}. The idea is simply
to use the cork $A$ and add some Casson handle $CH$ to it. The interior
of this construction is an exotic $\mathbb{R}^{4}$. Therefore we
have to consider the Casson handle and its construction in more detail.
Briefly, a Casson handle $CH$ is the result of attempts to embed
a disk $D^{2}$ into a 4-manifold. In most cases this attempt fails
and Casson \cite{Cas:73} looked for a substitute, which is now called
a Casson handle. Freedman \cite{Fre:82} showed that every Casson
handle $CH$ is homeomorphic to the open 2-handle $D^{2}\times\mathbb{R}^{2}$
but in nearly all cases it is not diffeomorphic to the standard handle
\cite{Gom:84,Gom:89}. The Casson handle is built by iteration, starting
from an immersed disk in some 4-manifold $M$, i.e. a map $D^{2}\to M$
with injective differential. Every immersion $D^{2}\to M$ is an embedding
except on a countable set of points, the double points. One can kill
one double point by immersing another disk into that point. These
disks form the first stage of the Casson handle. By iteration one
can produce the other stages. Finally consider not the immersed disk
but rather a tubular neighborhood $D^{2}\times D^{2}$ of the immersed
disk, called a kinky handle, including each stage. The union of all
neighborhoods of all stages is the Casson handle $CH$. So, there
are two input data involved with the construction of a $CH$: the
number of double points in each stage and their orientation $\pm$.
Thus we can visualize the Casson handle $CH$ by a tree: the root
is the immersion $D^{2}\to M$ with $k$ double points, the first
stage forms the next level of the tree with $k$ vertices connected
with the root by edges etc. The edges are evaluated using the orientation
$\pm$. Every Casson handle can be represented by such an infinite
tree. 

The main idea is the construction of a grope, an infinite union of
surfaces with non-vanishing genus, from the Casson handle. But the
grope can be represented by a sequence of polygons in the two-dimensional
hyperbolic space $\mathbb{H}^{2}$. This sequence of polygons is replaced
by one polygon with the same area. From this polygon we can construct
a codimension-one foliation on the 3-sphere as done by Thurston \cite{Thu:72}.
This 3-sphere is part of the boundary $\partial A$ of the Akbulut
cork $A$. Furthermore one can show that the codimension-one foliation
of the 3-sphere induces a codimension-one foliation of $\partial A$
so that the area of the corresponding polygons agree. 

Thus we are able to obtain a relation between an exotic $\mathbb{R}^{4}$
(of Bizaca as constructed from the failure of the smooth h-cobordism
theorem) and codimension-one foliation of the $S^{3}$. Two non-diffeomorphic
exotic $\mathbb{R}^{4}$implying non-cobordant codimension-one foliations
of the 3-sphere described by the Godbillon-Vey class in $H^{3}(S^{3},\mathbb{R})$
(proportional to the area of the polygon). This relation is very strict,
i.e. if we change the Casson handle then we must change the polygon.
But that changes the foliation and vice verse. Finally we obtained
the result:\\
\emph{The exotic $\mathbb{R}^{4}$ (of Bizaca) is determined by the
codimension-1 foliations with non-vanishing Godbillon-Vey class in
$H^{3}(S^{3},\mathbb{R}^{3})$ of a 3-sphere seen as submanifold $S^{3}\subset\mathbb{R}^{4}$.
We say: the exoticness is localized at a 3-sphere inside the small
exotic $\mathbb{R}^{4}$.}

\section{Geometry of string backgrounds and exotic $\mathbb{R}^{4}$}

In this section we take the point of view that exotic smoothness of
some small exotic $\mathbb{R}^{4}$'s when localized on $S^{3}\subset\mathbb{R}^{4}$,
correspond to some string geometry given by so-called $B$-fields
on $S^{3}$. The localization is understood as the representation
of the exotics by 3-rd integral or real cohomologies of $S^{3}$.
This correspondence is restricted to the classical limit of the geometry
of string backgrounds seen as a curved Riemannian manifold with B-field.
One can say that the small exotic smooth $\mathbb{R}^{4}$ given by
a localized $S^{3}$ is described by string geometry of $B$-fields
on this $S^{3}$. The correspondence can be extended to the string
regime of finite volume of $SU(2)$ WZW model.

\subsection{$SU(2)$ WZW model, D-branes and exotic $\mathbb{R}^{4}$}

In this subsection we want to focus on the change of smooth structure
on $\mathbb{R}^{4}$. As explained above, we realize the plan by considering
the changes as localized on $S^{3}$. Following \cite{AsselmeyerKrol2009,AsselmeyerKrol2009a},
this change gives rise to stringy effects, since the changes can be
described by computations in some 2D CFT, namely WZW models on $SU(2)$
at finite level.

\subsubsection{The dynamics of branes in the bosonic $SU(2)$ WZW model}

First we start with a discussion of the bosonic $SU(2)$ WZW model
and dynamics of branes in it. We deal here with $S^{3}$, i.e. the
metric of string background has non-zero curvature. In general, a
non-vanishing curvature $R(g)$ w.r.t. a non-constant metric $g$
of the background manifold $(M,g)$ on which bosonic string theory
is formulated, enforces the $H$ -field on $M$ to have a non-zero
value. This fact can be seen by using the string field equations(see
e.g. \cite{Schomerus2002}) \begin{equation}
R_{\mu\nu}(g)-\frac{1}{4}H_{\mu\rho\sigma}{H_{\nu}}^{\rho\sigma}={\cal O(\alpha')}\label{eq:R-H}\end{equation}
where $H=dB$ is the NSNS 3-form, $B=B(x)dx^{\mu}\wedge dx^{\nu}$
is the B-field, and dilaton field has a fixed, constant value. Furthermore
in the case of superstring theory this equation still holds true provided
all RR background fields vanish \cite{Schomerus2002}.

D-branes in group manifold $SU(2)$ (at the semi-classical limit)
are determined by wrapping the conjugacy classes of $SU(2)$, i.e.
(degenerated) 2-spheres described by a 2-sphere $S^{2}$ having two
poles each localized at a point. Due to the quantization conditions
there are $k+1$ D-branes on the level $k$ $SU(2)$ WZW model \cite{Schomerus2000,Schomerus2002,Alekseev1999}.
To grasp the dynamics of the branes one should deal with the gauge
theory on the stack of $N$ D-branes on $S^{3}$, quite similar to
the flat space case where noncommutative gauge theory emerges \cite{Alekseev1999b}.
Following this idea, the action for the brane dynamics is given by
the following argumentation.

Given $N$ branes of type $J$ (on top of each other), where $J$
is the representation of $SU(2)_{k}$ i.e. $J=0,\frac{1}{2},1,\,...\,,\frac{k}{2}$.
Then the dynamics of these branes (see \cite{Watamura2000}) is described
by the noncommutative action:

\begin{equation}
S_{N,J}=S_{YM}+S_{CS}=\frac{\pi^{2}}{k^{2}(2J+1)N}\left(\frac{1}{4}{\rm tr}(F_{\mu\nu}F^{\mu\nu})-\frac{i}{2}{\rm tr}(f^{\mu\nu\rho}{\rm CS}_{\mu\nu\rho})\right)\:.\label{eq:NoncommAction}\end{equation}
Here the curvature form $F_{\mu\nu}(A)=iL_{\mu}A_{\nu}-iL_{\nu}A_{\mu}+i[A_{\mu},A_{\nu}]+f_{\mu\nu\rho}A^{\rho}$
and the noncommutative Chern-Simons action reads ${\rm CS}_{\mu\nu\rho}(A)=L_{\mu}A_{\nu}A_{\rho}+\frac{1}{3}A_{\mu}[A_{\nu},A_{\rho}]$.
The fields $A_{\mu},\,\mu=1,2,3$ are defined on a fuzzy 2-sphere
$S_{J}^{2}$ and should be considered as $N\times N$ matrix-valued,
i.e. $A_{\mu}=\sum_{j,a}{\rm a}_{j,a}^{\mu}Y_{a}^{j}$ where $Y_{a}^{j}$
are fuzzy spherical harmonics and ${\rm a}_{j,a}^{\mu}$ are Chan-Paton
matrix-valued coefficients. $L_{\mu}$ are generators of the rotations
on fuzzy 2-spheres and they act only on fuzzy spherical harmonics
\cite{Schomerus2002}. The noncommutative action $S_{YM}$ was derived
from Connes spectral triples of the noncommutative geometry and was
aimed to describe Maxwell theory on fuzzy spheres \cite{Watamura2000}.
The equations of motion derived from the stationery points of (\ref{eq:NoncommAction}),
read\begin{equation}
L_{\mu}F^{\mu\nu}+[A_{\mu},F^{\mu\nu}]=0\:.\label{eq:eof}\end{equation}
The solutions of (\ref{eq:eof}) describe the dynamics of the branes,
i.e. the condensation processes on the brane configuration $(N,J)$
which results in another configuration $(N',J')$. A special class
of solutions, in the semi-classical $k\to\infty$ limit, can be obtained
from the $N(2J+1)$ dimensional representations of the algebra ${\rm su}(2)$.
For $J=0$ one has $N$ branes of type $J=0$, i.e. $N$ point-like
branes in $S^{3}$ at the identity of the group. Given another solution
corresponding to $J_{N}=\frac{N-1}{2}$, then one can show: this solution
can be firstly interpreted as the brane wrapping around the $S_{J_{N}}^{2}$
sphere but also as the condensed state of $N$ point-like branes at
the identity of $SU(2)$ \cite{Schomerus2002}:

\begin{equation}
(N,J)=(N,0)\to(1,\frac{N-1}{2})=(N',J')\label{eq:semi-class:N-to-N1}\end{equation}

Turning to the finite $k$ string regime of the $SU(2)$ WZW model
one can make use of the techniques of the boundary CFT when applied
to the analysis of Kondo effect \cite{Schomerus2002}. It follows
that there exists at the level of partition functions a continuous
shift between $N\chi_{j}(q)$ and the interfered sum of characters
$\sum_{j}N_{J_{N}j}^{\; l}\chi_{l}(q)$ where $N=2J_{N}+1$ (in the
vanishing value of the coupling constant) and $N_{J_{N}j}^{\; l}$
are Verlinde fusion rule coefficients. In the case of $N$ point-like
branes one can determine the decay product of these by considering
open strings ending on the branes. The result on the partition function
is

\[
Z_{(N,0)}(q)=N^{2}\chi_{0}(q)\]
which is continuously shifted to $N\chi_{J_{N}}(q)$ and next to $\sum_{j}N_{J_{N}J_{N}}^{\;\; j}\chi_{j}(q)$.
As the result we have the decay process \cite{Schomerus2002}

\begin{equation}\label{eq:stringN-to-N1}
\begin{array}{c}
Z_{(N,0)}(q)\to Z_{(1,J_{N})}\\[4pt]

(N,0)\to(1,J_{N})
\end{array}
\end{equation}which extends the similar process derived at the semi-classical $k\to\infty$
limit in the effective gauge theory (\ref{eq:semi-class:N-to-N1}),
however the representations $2J_{N}$ are bounded now, from the above,
by $k$.

\subsubsection{Brane charges and exotic $\mathbb{R}^{4}$}

Given the above dynamics of branes in the WZW $SU(2)$ model at string
regime, one can address the question of brane charges in a direct
way. This is based on the decay rule (\ref{eq:stringN-to-N1}) in
the supersymmetric WZW $SU(2)$ model. In this case we have a shift
of the level by $k\to k+2$ which measures the units of the NSNS flux
through $SU(2)=S^{3}$. One can see the supersymmetric model as strings
moving on $SU(2)$ with $k+2$ units of NSNS flux. From the CFT point
of view there exist currents $J^{a}$ which satisfy $k+2$ level of
the Kac-Moody algebra and free fermionic fields $\psi^{a}$ in the
adjoint representation of $su(2)$. However it is possible to redefine
the bosonic currents as

\[
J^{a}+\frac{i}{k}f_{\: bc}^{a}\psi^{b}\psi^{c}\]
which fulfill the current algebra commutation relation at the level
$k$. Here $f_{\: bc}^{a}$ are the structure constants of $su(2)$.
The fields $\psi^{a}$ commute with these currents. Thus we have the
splitting of the supersymmetric WZW $SU(2)$ model at level $k+2$
as WZW $SU(2)$ model at level $k$ times the theory of free fermionic
fields. 

Thus there are $k+1$ stable branes wrapping the conjugacy classes
numbered by $J=0,\frac{1}{2},...,\frac{k}{2}$. The decaying process
(\ref{eq:stringN-to-N1}) says that placing $N$ point-like branes
(each charged by the unit $1$) at the pole $e$ they can decay to
the spherical brane $J_{N}$ wrapping the conjugacy class. Taking
more point-like branes to the stack at $e$ gives the more distant
$S^{2}$ branes until reaching the opposite pole $-e$ where we have
single point-like brane with the opposite charge $-1$. Having identify
$k+1$ units of the charge with $-1$ we arrive at the conclusion
that the group of charges is $\mathbb{Z}_{k+2}$. More generally the
charges of branes on the background $X$ with non-vanishing $H\in H^{3}(X,\mathbb{Z})$
are described by the twisted $K$ group $K_{H}^{\star}(X)$ (see e.g.
\cite{MathaiMurray2001}). In the case of $SU(2)$ we get the group
of RR charges as above for $K=k+2$ \begin{equation}
K_{H}^{\star}(S^{3})=\mathbb{Z}_{K}\end{equation}

Turning to the exotic $\mathbb{R}^{4}$ case based on \cite{AsselmeyerKrol2009},
we have for a given nonzero integral class $H\in H^{3}(S^{3},\mathbb{Z})$
exotic $\mathbb{R}_{H}^{4}$. And conversely, given this exotic $\mathbb{R}_{H}^{4}$
we recover the class $H\in H^{3}(S^{3},\mathbb{Z})$. Certain topological
conditions have to be fulfilled: the 3-sphere, i.e. $SU(2)$, is seen
as a part of the boundary of the Akbulut cork. It is the attachment
of the Casson handle to the Akbulut cork, which determine the exotic
$\mathbb{R}_{H}^{4}$. Thus we can correlate ambient exotic smoothness
of $\mathbb{R}_{H}^{4}$ with the classes $H\in H^{3}(S^{3},\mathbb{Z})$
provided $S^{3}$ is the part of the boundary of the Akbulut cork.
Moreover, in \cite{AsselmeyerKrol2009a} was shown that exotic smooth
$\mathbb{R}_{H}^{4}$ deforms K-theory $K(S^{3})$ toward equivariant
one $K_{H}(S^{3})$. Thus, we obtain the following important observation:\emph{
certain small exotic $\mathbb{R}^{4}$'s generate the group of RR
charges of D-branes in the curved background of $S^{3}\subset\mathbb{R}^{4}$.} 

Then we arrive at the correspondence:

\begin{theorem}

The classification of RR charges of the branes on the background given
by the group manifold $SU(2)$ at the level $k$ (hence the dynamics
of D-branes in $S^{3}$ in stringy regime) is correlated with the
exotic smoothness on $\mathbb{R}^{4}$ containing this $S^{3}=SU(2)$
as the part of the boundary of the Akbulut cork. 

\end{theorem} 

We can give yet another interpretation of the 4-exoticness which appears
on flat $\mathbb{R}^{4}$ in this context. Exotic smoothness of $\mathbb{R}^{4}$,
$\mathbb{R}_{H}^{4}$, determines the collection of stable D-branes
on $SU(2)$ at the level $k$ of the WZW model, where $k=[H]\in H^{3}(S^{3},\mathbb{Z})$.
Thus, the string-finite $k$-level of the WZW model characterizes
exotic 4-smoothness. Recall that in the case of $H=0$ (e.g. $B$
constant in a flat space, i.e. in $k\to\infty$ limit) the smooth
structure on $\mathbb{R}^{4}$ is the standard one \cite{AsselmeyerKrol2009}.
Thus the exotic smoothness on $\mathbb{R}^{4}$ translates the 4-curvature
to the non-zero H-field on $S^{3}$ of finite volume in string units.
This is similar to the effect of string field equations relating $R$
and $H$ as in (\ref{eq:R-H}), though it holds now between different
spaces ($\mathbb{R}^{4}$ and $S^{3}$).

\subsection{$SU(2)$ WZW model in the geometry of the stack of NS5-branes}

The manifold $SU(2)=S^{3}$ is the only group manifold which became
relevant so far for the description of small exotic $\mathbb{R}^{4}$.
From the other side it is the only one which appears directly as part
of a string background (namely one generated by NS5-branes). That
is why the connection of 4-exotics and string theory can be naturally
formulated in the geometry of the stack of NS5-branes. Let us briefly
describe this string background \cite{Schomerus2000,Schomerus2002,Bachas2000}.

We consider a configuration of $k$ coincident supersymmetric NS5-branes
in type II theory. The full fivebrane background is (in string frame) 

\begin{equation}\label{eq:NS5-background}
\begin{array}{c}
ds^{2}=dx^{2}+f(r)dy^{2}\\[4pt]

e^{2\phi}=g_{s}^{2}f(r)\\[4pt]

f(r)=1+\frac{k\alpha'}{r^{2}} \\[4pt]

H_{IJK}=k\alpha'\epsilon_{IJK}
\end{array}
\end{equation}where $x$ are the $5+1$ longitudinal coordinates along NS5-branes
referred to by indices $\mu$, $\nu$, etc., $y$ being 4 transverse
coordinates referred to by indices $I$, $J$, $K$ ... and $r=|y|$,
$1/\alpha'\sim$ string tension. The fields of this background are
given by

\begin{equation}
\begin{array}{c}
e^{2\Phi}=1+\sum_{j=1}^{k}\frac{l_{s}^{2}}{|y-y_{j}|^2}\\[4pt]

g_{IJ}=e^{2\Phi}\delta_{IJ}\\[4pt]

g_{\mu\nu}=\eta_{\mu\nu} \\[4pt]

H_{IJK}=-\epsilon_{IJKL}\partial^{L}\Phi
\end{array}
\end{equation}where $y_{j},j=1,...,k$ are the positions of the NS5-branes. When
the branes coincide at 0, $y_{j}=0$, the near horizon solutions $y\to0$,
are

\begin{equation}
\begin{array}{c}
e^{2\Phi}=\frac{kl_{s}^{2}}{|y|^{2}}\\[4pt]

g_{IJ}=e^{2\Phi}\delta_{IJ}\\[4pt]

g_{\mu\nu}=\eta_{\mu\nu} \\[4pt]

H_{IJK}=-\epsilon_{IJKL}\partial^{L}\Phi
\end{array}
\end{equation}

In the near-horizon limit $r=|y|^{2}\to0$, the background factorizes
into a radial component as well in a $S^{3}$ and flat 6-dimensional
Minkowski spacetime. Strings propagating at this limiting background
are described by the exact world-sheet CFT with the target $\mathbb{R}^{5,1}\times\mathbb{R}_{\phi}\times S_{k}^{3}$.
Here $\mathbb{R}_{\phi}$ is the real line with the parameter $\phi$
which is a scalar corresponding to the ,,linear dilaton'' 

\begin{equation}
\begin{array}{c}
\Phi=-\sqrt{\frac{1}{2k}}\phi\\[4pt]

\phi=\sqrt{\frac{k}{2}}\log\frac{r}{kl_{s}^{2}}

\end{array}
\end{equation}

The flat Minkowski space $\mathbb{R}^{5,1}$ is longitudinal to the
directions of NS5-branes, $S_{k}^{3}$ is $SU(2)_{k}$ and is a level
$k$ WZW supersymmetric CFT (SCFT) on $SU(2)$ as discussed in the
previous section. This $S^{3}$ corresponds to the angular coordinates
of the transversal $\mathbb{R}^{4}$. We see that the infinite ,,throat''
$\mathbb{R}_{\phi}\times S_{k}^{3}$, emerges with the metric of the
background (in the string frame)

\[
ds^{2}=dx_{6}^{2}+d\phi^{2}+kl_{s}d\Omega_{3}^{2}\,,\: g_{s}^{2}(\phi)=e^{-2\phi/\sqrt{k}l_{s}}\:.\]
This background is obtained in the near horizon geometry (i.e. $\phi\to-\infty$
($r\to0$) ) of the stack of $k-2$ NS5-branes in type II string theory
and is in fact a SCFT on the throat. The NS5-branes are placed at
$\phi\to-\infty$ and string theory is in a strong-coupling regime,
i.e. $g_{s}\sim\exp(2\Phi)$. In the opposite limit ($\phi\to+\infty$,
or $r\to+\infty$) gives asymptotically flat 10-manifold and string
theory is weakly coupled in that limit. This is essentially the CHS
(Callan, Harvey, Strominger \cite{CHS1991}) exact string theory background
where $SU(2)$ WZW model appears at suitable level $k$. 

Given the CHS limiting geometry of $N$ NS5-branes we have the 4-dimensional
tube $\mathbb{R}_{\phi}\times S^{3}$. The volume of $S^{3}$ in string
units is finite and correlated with the number of NS5-branes by $N=k-2$
\cite{Bachas2000}. We take an exotic $\mathbb{R}_{H}^{4}$ for $[H]=k[\:]\in H^{3}(S^{3},\mathbb{Z})$.
This can be achieved more directly by considering the Akbulut cork
$A_{H}$ with the boundary, $\partial A_{H}=\Sigma_{H}$, a homology
3-sphere. As was shown in \cite{AsselmeyerKrol2009} $\Sigma_{H}$
contains $S^{3}$ such that the codimension-1 foliations of it generates
the foliations of $\Sigma_{H}$. Expressed in a different way, this
foliation is generated by the Casson handles attached to $A$. Thus
the Akbulut cork attached to the Casson handle (or an open neighborhood
$N(A)$ of the cork $A$) determines the small exotic smoothness of
$\mathbb{R}_{H}^{4}$ \cite{GomSti:1999,AsselmeyerKrol2009}. Two
different exotic smoothness structures on the $\mathbb{R}^{4}$ are
given by two non-cobordant codimension-one foliations of the 3-sphere.
Moreover, the cobordism classes of codimension-1 foliations of $S^{3}$
are classified by the Godbillon-Vey invariants which are elements
of $H^{3}(S^{3},\mathbb{R})$. In our case we deal with integral 3-rd
cohomologies $[H]\in H^{3}(S^{3},\mathbb{Z})$. Thus, the embedding
of the Akbulut cork (determined by the Casson handle) in the ambient
$\mathbb{R}^{4}$ is determined by the integral classes $k[\:]\in H^{3}(S^{3},\mathbb{Z})$.
By using the diffeomorphism $\Sigma_{H}=\Sigma_{H}\#S^{3}$ we obtain
a $S^{3}$ as part of the boundary $\Sigma_{H}$ of the Akbulut cork.
By using the identification $S^{3}=SU(2)$ in the context of the string
background of $N$ NS5-branes we have the following result:

\begin{theorem}\label{Th:Ns5Branes}

In the geometry of the stack of NS5-branes in type II superstring
theories, adding or subtracting a NS5-brane is correlated with the
change of the smoothness structure on the transversal $\mathbb{R}^{4}$. 

\end{theorem}

Now the geometry of the tube $\mathbb{R}_{\phi}\times S^{3}$ is defined
with respect to the ambient standard $\mathbb{R}^{4}$. Interpreting
the $S^{3}$ as part of the boundary of the Akbulut cork for some
exotic smooth $\mathbb{R}_{H}^{4}$, the factor $S_{k}^{3}$ in the
background is correlated with $\mathbb{R}_{H}^{4}$, $[H]=k[\:]\in H^{3}(S^{3},\mathbb{Z})$.
Changing $k$ causes a change of smoothness for the ambient $\mathbb{R}^{4}$.
As explained above, this smoothness is determined by the embedding
of the Akbulut cork. However, the change of the smoothness of the
transversal $\mathbb{R}^{4}$ affects the geometry of the tube. Thus
the background $\mathbb{R}^{5,1}\times\mathbb{R}_{\phi}\times SU(2)_{k}$
is correlated by the above topological arguments with another smooth
geometry, namely $\mathbb{R}^{5,1}\times\mathbb{R}_{H}^{4}$ where
we now interpret $\mathbb{R}_{H}^{4}$ as an exotic $\mathbb{R}^{4}$
transversal to $\mathbb{R}^{5,1}$. This is precisely the geometry
which is sensitive to the number of NS5-branes as in Theorem \ref{Th:Ns5Branes}.
Thus string theory with CHS limiting geometry deals with the geometry
of $\mathbb{R}^{5,1}\times\mathbb{R}_{H}^{4}$. However, possible
correction terms in the DBI action for branes in such backgrounds
can appear. 

The construction of any smooth metric on an exotic $\mathbb{R}_{H}^{4}$
in explicit form is a very complicated mathematical task at present.
Nevertheless string theory touches these intractable geometries and
reflects their effects some branes configurations in the CHS limit.
Thus string theory gives information relating exotic $\mathbb{R}^{4}$
regions of the background. Especially one should keep the following
fact in mind: by working with an exotic $\mathbb{R}_{H}^{4}$ one
has to forget the factorization like in the smooth CHS geometry, i.e.
$\mathbb{R}_{\phi}\times S^{3}$. This fact can be understood by considering
the end of $\mathbb{R}^{4}$, i.e. this part of $\mathbb{R}^{4}$
which extends to infinity%
\footnote{An end of a space $X$ is the limit of a sequence $U_{1}\subset U_{2}\subset\ldots$
of complements $U_{n}=X\setminus K_{n}$ where $K_{1}\subset K_{2}\subset\ldots$
is an ascending sequence of compact sets $K_{n}$ whose interiors
cover $X$.%
}. The end of the $\mathbb{R}^{4}$ is $S^{3}\times\mathbb{R}$. Then
an exotic $\mathbb{R}^{4}$ has an exotic end $(S^{3}\times\mathbb{R})_{\theta}$.
But $(S^{3}\times\mathbb{R})_{\theta}$ do not factorize (or is globally
foliated by the leafs $S^{3}\times\left\{ t\right\} $ for all $t\in\mathbb{R}$)
\cite{Fre:79} otherwise it is the end of the standard smooth $\mathbb{R}^{4}$.
Thus we have the following correspondence: \emph{The change of exotic
smoothness of $\mathbb{R}_{H}^{4}$ in }$\mathbb{R}^{5,1}\times\mathbb{R}_{H}^{4}$\emph{
to the standard one, gives the factorization of the CHS limiting geometry.
String theory in this limiting CHS geometry remembers from which $\mathbb{R}_{H}^{4}$
it was projected by the level $k$ of $SU(2)_{k}$ factor.} Thus the
information about the string background $\mathbb{R}^{5,1}\times\mathbb{R}_{\phi}\times SU(2)_{k}$
is originally encoded purely geometrically as $\mathbb{R}^{5,1}\times\mathbb{R}_{H}^{4}$.
To understand precisely this relation in analytical terms is possible.
But we will present it in a separate paper, since this requires some
mathematical work on exotic \emph{$\mathbb{R}_{H}^{4}$.}

In that way we could have a theory producing some exact results of
string theory. This theory would have an (exotic) 4-geometry (given
by a foliation) on the Euclidean space $\mathbb{R}^{4}$ as the fundamental
structure. The product factorization of the end $S^{3}\times\mathbb{R}$
gives the standard smoothness and also the background of the string
theory. Moreover, these (small) exotic $\mathbb{R}_{H}^{4}$ are all
embeddable in standard $\mathbb{R}^{4}$ and carry string theory information
about the exact string backgrounds and brane charges. Then we will
obtain a full 4-dimensional description: exotic $\mathbb{R}_{H}^{4}$
embedded in $\mathbb{R}^{4}$ and projecting to the CHS string background
of $k$ NS5-branes where $H=k[\:]\in H^{3}(S^{3},\mathbb{Z})$. One
goal of our future work is the collecting of arguments that such a
theory could be considered as more fundamental than string theory
itself. Instead we will propose here a general heuristic rule in our
geometrical setting:

R1. \emph{Lifted D-branes probing the exotic geometry }$\mathbb{R}^{5,1}\times\mathbb{R}_{H}^{4}$\emph{
are projected on to D-branes of type II string theory probing the
factorized geometry, $\mathbb{R}^{5,1}\times\mathbb{R}_{\phi}\times SU(2)_{k}$
based on standard smooth product $\mathbb{R}_{\phi}\times S^{3}$.
Again, the dependence of the calculations on $k$ is a direct consequence
of the exoticness of }$\mathbb{R}_{H}^{4}$.\emph{ The projection
is driven by the factorization of $smooth$ $\mathbb{R}^{4}$ such
that the topological product of axes becomes smooth one.} 

,,Lifted D-branes probing the exotic geometry $\mathbb{R}^{5,1}\times\mathbb{R}_{H}^{4}$''
serves here merely as a description of the correspondence following
Theorem \ref{Th:Ns5Branes}. Furthermore we do not formulate any result
depending on the actual existence of these D-branes. What is their
exact status (in the lifted theory) will become evident when modified
DBI action and solutions are presented.

Rule R1 is based on the observation in \cite{AsselmeyerKrol2009}
that various nonstandard smoothings of $\mathbb{R}^{4}$ can be grasped
by the effects of $H^{3}(S^{3},\mathbb{Z})$ and that the factorization
$\mathbb{R}_{\phi}\times S^{3}$ of the end gives the standard smooth
$\mathbb{R}^{4}$. Following this rule we can consider many examples
of D-branes in the above background (see e.g. \cite{GiveonAntoniadis2000,GiveonKutasov2000,YunKwon2009,Ribault2003,ChenSun2005}),
as referring to 4-exoticness. Let us discuss briefly the case of little
string theory (LST) in this context. 

Type II string theory on $\mathbb{R}^{5,1}\times\mathbb{R}_{\phi}\times SU(2)_{k}$
is given by the SCFT on the infinite ,,throat'' of the background,
i.e. $\mathbb{R}_{\phi}\times S^{3}$. Then this theory was proposed
to be approachable via holography by using duality. The holographically
dual theory is the 6-dimensional \emph{little string theory} \cite{GiveonKutasov2000,Aharony2002}.
LST has possible experimental signatures at the TeV scales after the
compactification on the torus \cite{GiveonAntoniadis2000}. 

LST's are non-local theories without gravity and can be described
in the limit $g_{s}\to0$ in the theory of $k$ NS5-branes. In this
limit the bulk degrees of freedom decouple, hence gravity does. This
6-dimensional LST without gravity is holographically dual to the type
II string theory formulated on the background $\mathbb{R}^{5,1}\times\mathbb{R}_{\phi}\times SU(2)_{k}$
\cite{GiveonAntoniadis2000}. From the rule R1 above, it follows that
LST refers also to exotic $\mathbb{R}_{H}^{4}$'s. However the perturbative
calculations are hardly performed in LST since the string coupling
$g_{s}$ diverges in the dual string background along the tube, and
LST is sensitive on this background. One usually regulates the geometry
via chopping the tube. But the decomposition of the SCFT $SU(2)_{k}$
on $S_{Y}^{1}\times SU(2)_{k}/U(1)$ can be performed. Here $SU(2)_{k}/U(1)$
is the minimal $N=2$ model at the level $k$ and $S_{Y}^{1}$ is
the Cartan subalgebra of $SU(2)$ with the parameter $Y$. The dependence
on $k$ is crucial in this reformulation since it refers to 4-exotics
by theorem \ref{Th:Ns5Branes} and the rule R1. Thus we have the SCFT
$\mathbb{R}_{\phi}\times S_{Y}^{1}\times\frac{SU(2)_{k}}{U(1)}$ instead
of the tube $\mathbb{R}_{\phi}\times SU(2)_{k}$. The chopping of
the strong coupling region is now performed by taking the SCFT $\frac{SL(2)_{k}}{U(1)}$
instead of $\mathbb{R}_{\phi}\times S_{Y}^{1}$ which means replacing
the background $\mathbb{R}^{5,1}\times\mathbb{R}_{\phi}\times SU(2)_{k}$
by $\mathbb{R}^{5,1}\times\frac{SL(2)_{k}}{U(1)}\times\frac{SU(2)_{k}}{U(1)}$.
This means, on the level of $k$ NS5-branes, the separation of these
5-branes along the transverse circle of radius $L$. Now the double-scale
limit of LST is the one when taking both $g_{s}$ and $L$ to zero
while $\frac{L}{g_{s}}$ remains constant. 

Following \cite{GiveonKutasov2000} we can take systems of D4, D6-branes
between separated NS5-branes. The various expressions like correlation
functions can be now calculated perturbatively in the holographically
dual 6-dimensional LST theory. Besides, suitable compactifications
may refer to the spectra with the TeV scale of the standard model
of particles. The dependence on $k$ of some of these expressions
can be seen again as the signature of that these expressions were
obtained by standard factorization and projection from non-factorized
exotic 4-geometries.

\emph{Exoticness of the 4-space transversal to the world-volume of
NS5-branes, is reflected in specific perturbative spectra of D-branes
when calculated in dual 6-dimensional LST. When compactifying this
LST on 2 directions longitudinal to the 5-brane one gets spectra which
could be sensitive on transversal exoticness of $\mathbb{R}^{4}$. }

This is in fact the reformulation of the rule R1. The NS5-branes backgrounds
show that string theory computations ,,feel'' the 4-exoticness. On
the ather hand, these 4-exotic regions contain information about branes
in certain string backgrounds.

\section{Discussion and conclusions}

In this paper we tried to give a partial answer to the important question:
Is it possible that string theory deals with 4-dimensional structures
directly neither by implementing compactifications nor by phenomenological
models-building, and these structures would have a physical meaning?

We propose that the structures are nonstandard smoothings of the Euclidean
4-space. Here we present the scenario where exotic smoothness on Euclidean
$\mathbb{R}^{4}$ appears in string theory and is correlated with
the charges, hence dynamics, of NS and D branes in certain string
backgrounds. We gave topological reason for the above correspondence:
when a WZW model on $SU(2)$ at the level $k$ appears in string theory
as a part of exact string backgrounds then we will get a correspondence
by placing this $SU(2)=S^{3}$ as a part of the boundary of the Akbulut
cork. Thus exotic smoothness of the Euclidean $\mathbb{R}_{k}^{4}$
is determined by the 3-sphere localized at the boundary of the Akbulut
cork. In the CHS limiting geometry of the stack of $k$ NS5-branes
we are able to show that there exists a 4-region of this solution
which is projected from the unique exotic $\mathbb{R}_{k}^{4}$ via
factorization. In that way the dependence on $k$ appears and the
string solution remembers the original exoticness via this dependence.
We conjecture, that a theory without projection should be considered
as fundamental for string theory. In the CHS background the lifted
theory describes small exotic $\mathbb{R}^{4}$'s as embedded in the
standard $\mathbb{R}^{4}$. These data are projected on the factorized
string background $\mathbb{R}^{5,1}\times\mathbb{R}_{\phi}\times SU(2)_{k}.$
Moreover, the close relation with string theory opens the possibility
that exotic metrics on $\mathbb{R}_{k}^{4}$ could be derived from
string theory calculations. 

At present the construction of any metric for any exotic $\mathbb{R}^{4}$
is untractable. To recognize 4-exotic geometry from the point of view
of string theory one could try to formulate a modified DBI action
for D-branes in the lifted theory which would explore exotic 4-spaces,
similarly as various D-branes do for transversal $\mathbb{R}_{\phi}\times SU(2)_{k}$.
One possibility is the usage of D3-branes in this string background.
These D3-branes fill the 3-space $\mathbb{R}_{\phi}\times S^{2}$
where 2-spheres are the conjugacy classes of $SU(2)$. Again, the
number of allowed conjugacy classes depend on the level $k$ WZW model
on $SU(2)$ whereas the standard smooth structure on $\mathbb{R}_{\phi}\times SU(2)$
follows from this factorization. It was proposed in \cite{AsselmeyerKrol2009}
to consider some correlation functions between states in the WZW model,
and use these states to characterize exotic 4-smoothness. One could
also deal with a kind of superposition referring to quantum correlation
functions. Or, one should consider wild embeddings of $S^{2}$ in
$\mathbb{R}_{\phi}\times S^{2}$ which directly refers to exotic $\mathbb{R}^{4}$'s
\cite{AsselmeyerKrol2009}. Then one loses the factorization, and,
on the other hand, approaches the branes in string theory as a quantum
object. These ideas will be the topic of our next paper.

Further, one could wonder whether exotic smooth $\mathbb{R}^{4}$'s
are suitable objects when dealing with 4-dimensional physics: Are
these structures relevant to physics at all? This important question
was already answered, though partially in some research papers \cite{BraRan:93,Bra:94a,Bra:94b,Ass:96,AssBra:2002,Ass2010}
and see also the textbook \cite{Asselmeyer2007}. In particular in
\cite{AsselmeyerKrol2009} we showed that exotic smoothness of an
open 4-region in spacetime have the same effect as the existence of
magnetic monopoles, i.e. exotic smoothness induces the quantization
condition for the electric charge. By using \cite{AsselmeyerKrol2010},
one finds many further arguments to consider the exotic $\mathbb{R}^{4}$'s
as quantum object, i.e. the spacetime induces the quantization processes.
The work on uncovering 4-dimensional physics from exotic $\mathbb{R}^{4}$'s
by using the quantum-physical point of view is currently developed,
which should be seen as complimentary to the string theory thread.
Completing this work should give more essential understanding of the
formalism of string theory as referring to 4-dimensional physics which
is not covered by compactification. On the other hand our understanding
of the phenomenon of exotic 4-smoothness on open manifolds has a chance
to be broaden.

\section*{Acknowledgment}

T.A. wants to thank C.H. Brans and H. Ros\'e for numerous discussions
over the years about the relation of exotic smoothness to physics.
J.K. benefited much from the explanations given to him by Robert Gompf
regarding 4-smoothness several years ago, and discussions with Jan
S{\l}adkowski.

.

%\bibliographystyle{plain}
%\addcontentsline{toc}{section}{\refname}\bibliography{diffbib,foliation-gerbes,knots}

\begin{thebibliography}{10}

\bibitem{Aharony2002}
O.~Aharony.
\newblock The non-{ADS}/non-{CFT} correspondence, or three different paths to
  {QCD}.
\newblock {\em Lectures notes from the Carges 2002 summer school}, 2002.
\newblock arXiv:hep-th/0212193v2.

\bibitem{Akb:96}
S.~Akbulut.
\newblock Lectures on {S}eiberg-{W}itten invariants.
\newblock {\em Turkish J. Math.}, {\bf 20}:95--119, 1996.

\bibitem{Akbulut08}
S.~Akbulut and K.~Yasui.
\newblock Corks, plugs and exotic structures.
\newblock {\em Journal of Gokova Geometry Topology}, 2:40--82, 2008.
\newblock arXiv:0806.3010.

\bibitem{Akbulut09}
S.~Akbulut and K.~Yasui.
\newblock Knotted corks.
\newblock {\em J Topology}, 2:823--839, 2009.
\newblock arXiv:0812.5098.

\bibitem{Alekseev1999}
A.Y. Alekseev and V.~Schomerus.
\newblock D-branes in the {WZW} model.
\newblock {\em Phys. Rev. D}, 60:061901, 1999.

\bibitem{GiveonAntoniadis2000}
S.~Antoniadis, I.~Dimopoulos and A.~Giveon.
\newblock Little string theory at a {T}e{V}.
\newblock {\em JHEP}, 0105055, 2001.
\newblock arXiv:hep-th/0103033v2.

\bibitem{Ass:96}
T.~Asselmeyer.
\newblock Generation of source terms in general relativity by differential
  structures.
\newblock {\em Class. Quant. Grav.}, {\bf 14}:749 -- 758, 1996.

\bibitem{Ass2010}
T.~Asselmeyer-Maluga.
\newblock Exotic smoothness and quantum gravity.
\newblock {\em Class. Quantum Grav.}, 27:165002, 2010.
\newblock arXiv:1003.5506v1 [gr-qc].

\bibitem{AssBra:2002}
T.~Asselmeyer-Maluga and C.H. Brans.
\newblock Cosmological anomalies and exotic smoothness structures.
\newblock {\em Gen. Rel. Grav.}, 34:1767--1771, 2002.

\bibitem{Asselmeyer2007}
T.~Asselmeyer-Maluga and C.H. Brans.
\newblock {\em Exotic {S}moothness and {P}hysics}.
\newblock WorldScientific Publ., Singapore, 2007.

\bibitem{AsselmeyerKrol2009a}
T.~Asselmeyer-Maluga and J.~Kr{\'o}l.
\newblock Gerbes on orbifolds and exotic smooth {$R^4$}.
\newblock subm. to Comm. Math. Phys., arXiv: 0911.0271, 2009.

\bibitem{AsselmeyerKrol2009}
T.~Asselmeyer-Maluga and J.~Kr{\'o}l.
\newblock Gerbes, {SU(2)} {WZW} models and exotic smooth {$R^4$}.
\newblock submitted to Comm. Math. Phys., arXiv: 0904.1276, 2009.

\bibitem{AsselmeyerKrol2010}
T.~Asselmeyer-Maluga and J.~Kr{\'o}l.
\newblock Exotic smooth $\mathbb{R}^4$, noncommutative algebras and
  quantization.
\newblock submitted to Comm. Math. Phys., arXiv: 1001.0882, 2010.

\bibitem{Alekseev1999b}
V.~Schomerus A.Y.~Alekseev, A.~Recknagel.
\newblock Non-commutative world-volume geometries: Branes on {SU(2)} and fuzzy
  spheres.
\newblock {\em JHEP}, 9909:023, 1999.

\bibitem{Biz:94a}
Z.~Bizaca.
\newblock A handle decomposition of an exotic {${\mathbb R}^4$}.
\newblock {\em J. Diff. Geom.}, 39:491 -- 508, 1994.

\bibitem{BizGom:96}
{\u Z}.~Bi{\u z}aca and R~Gompf.
\newblock Elliptic surfaces and some simple exotic {${\Bbb {R}}^4$}'s.
\newblock {\em J. Diff. Geom.}, {\bf 43}:458--504, 1996.

\bibitem{MathaiMurray2001}
P.~Bouwknegt, A.L. Carey, V.~Mathai, M.K. Murray, and D.~Stevenson.
\newblock Twisted {K}-theory and {K}-theory of bundle gerbes.
\newblock {\em Comm. Math. Phys.}, 228:17--45, 2002.
\newblock arXiv: hep-th/0106194.

\bibitem{Bra:94b}
C.H. Brans.
\newblock Exotic smoothness and physics.
\newblock {\em J. Math. Phys.}, {\bf 35}:5494--5506, 1994.

\bibitem{Bra:94a}
C.H. Brans.
\newblock Localized exotic smoothness.
\newblock {\em Class. Quant. Grav.}, {\bf 11}:1785--1792, 1994.

\bibitem{BraRan:93}
C.H. Brans and D.~Randall.
\newblock Exotic differentiable structures and general relativity.
\newblock {\em Gen. Rel. Grav.}, {\bf 25}:205, 1993.

\bibitem{Bachas2000}
C.~Schweigert C.~Bachas, M.~Douglas.
\newblock Flux stabilization of {D}-branes.
\newblock {\em JHEP}, 0005:048, 2000.

\bibitem{CHS1991}
C.~Callan, J.~Harvey, and A.~Strominger.
\newblock Supersymmetric string solitons.
\newblock In {\em String Theory and Quantum Gravity}, Proceedings, page 208,
  Trieste 1991, 1991.
\newblock arXiv:hep-th/9112030.

\bibitem{Watamura2000}
J.~Carow-Watamura and S.~Watamura.
\newblock Noncommutative geometry and gauge theory on fuzzy sphere.
\newblock {\em Commun. Math. Phys.}, 212:395, 2000.
\newblock arXiv:hep-th/9801195.

\bibitem{Cas:73}
A.~Casson.
\newblock {\em Three lectures on new infinite constructions in 4-dimensional
  manifolds}, volume~62.
\newblock Birkh{\"a}user, progress in mathematics edition, 1986.
\newblock Notes by Lucian Guillou, first published 1973.

\bibitem{ChenSun2005}
B.~Chen and B.~Sun.
\newblock Note on {DBI} dynamics of {D}brane near {NS}5-branes.
\newblock {\em Phys. Rev. D}, 72:046005, 2005.
\newblock arXiv:hep-th/0501176.

\bibitem{CuFrHsSt:97}
C.~Curtis, M.~Freedman, W.-C. Hsiang, and R.~Stong.
\newblock A decomposition theorem for h-cobordant smooth simply connected
  compact 4-manifolds.
\newblock {\em Inv. Math.}, {\bf 123}:343--348, 1997.

\bibitem{DonKro:90}
S.~Donaldson and P.~Kronheimer.
\newblock {\em The Geometry of Four-Manifolds}.
\newblock Oxford Univ. Press, Oxford, 1990.

\bibitem{GiveonKutasov2000}
S.~Elitzur1, A.~Giveon, D.~Kutasov, E.~Rabinovici, and G.~Sarkissian.
\newblock D-branes in the background of {NS} fivebranes.
\newblock {\em JHEP}, 0008:046, 2000.
\newblock arXiv:hep-th/0005052.

\bibitem{Schomerus2000}
S.~Fredenhagen and V.~Schomerus.
\newblock Branes on group manifolds, gluon condensates, and twisted {K}-theory.
\newblock {\em JHEP}, 04:007, 2001.
\newblock arXiv:hep-th/0012164.

\bibitem{Fre:79}
M.H. Freedman.
\newblock A fake {$S^3\times R$}.
\newblock {\em Ann. of Math.}, {\bf 110}:177--201, 1979.

\bibitem{Fre:82}
M.H. Freedman.
\newblock The topology of four-dimensional manifolds.
\newblock {\em J. Diff. Geom.}, {\bf 17}:357 -- 454, 1982.

\bibitem{Gom:84}
R.~Gompf.
\newblock Infinite families of casson handles and topological disks.
\newblock {\em Topology}, {\bf 23}:395--400, 1984.

\bibitem{Gom:89}
R.~Gompf.
\newblock Periodic ends and knot concordance.
\newblock {\em Top. Appl.}, {\bf 32}:141--148, 1989.

\bibitem{GomSti:1999}
R.E. Gompf and A.I. Stipsicz.
\newblock {\em 4-manifolds and {K}irby {C}alculus}.
\newblock American Mathematical Society, 1999.

\bibitem{YunKwon2009}
Gyeong~Yun Jun and Pyung~Seong Kwon.
\newblock D-brane orbiting {NS}5-branes.
\newblock {\em JHEP}, 1001:062, 2009.
\newblock arXiv:0911.4557.

\bibitem{Krol2010}
J.~Kr{\'o}l.
\newblock ({Q}uantum) gravity effects via exotic $\mathbb{R}^4$.
\newblock {\em Ann. Phys. (Berlin)}, 19:No. 3--5, 355--358, 2010.

\bibitem{Ribault2003}
S.~Ribault.
\newblock D3-branes in {NS}-branes backgrounds.
\newblock {\em JHEP}, 0302, 2003.
\newblock arXiv:hep-th/0301092.

\bibitem{Schomerus2002}
V.~Schomerus.
\newblock Lectures on branes in curved backgrounds.
\newblock arXiv:hep-th/0209241, 2002.

\bibitem{Thu:72}
W.~Thurston.
\newblock Noncobordant foliations of {$S^3$}.
\newblock {\em BAMS}, 78:511 -- 514, 1972.

\end{thebibliography}

\end{document}